\documentclass[english,10pt]{article} 
\usepackage[T1]{fontenc} \usepackage[latin1]{inputenc}
\usepackage{babel} \usepackage{epsfig}
\usepackage{amsmath}
\usepackage{amsfonts}
\newcommand{\pde}[1]{\frac{\partial}{\partial {#1}}}
\newcommand{\pdz}[1]{\frac{\partial^2}{\partial{#1}^2}}
\newcommand{\tde}[1]{\frac{d}{d {#1}}}

\newcommand{\TT}{{\cal{T}}}
\newcommand{\PP}{{\cal{P}}}

\newcommand{\JJ}{{\cal{J}}}

\newcommand{\intrinsicl}{\mathfrak{L}}
\newcommand{\contvar}{{x}}
\newcommand{\mean}[1]{\left<{#1}\right>}

\newcommand{\pd}[2]{\frac{\partial^{#1}}{\partial{#2}^{#1}}}
\begin{document}

\title{Drift and Diffusion in Periodically Driven  Renewal Processes} 
\author{T.~Prager\footnote{e-mail tobias@physik.hu-berlin.de } and
  L.~Schimansky-Geier\\
 Institute of Physics, Humboldt-University at Berlin,\\ Newtonstr. 15, D-12489 Berlin, Germany} 
\date{\today}

\maketitle

\begin{abstract}
  We consider the drift and diffusion properties of periodically
  driven  renewal
  processes. These processes 
  are defined by a periodically time dependent waiting time
  distribution, which governs the interval between subsequent events.
  We show that the growth of the cumulants of the number of events is
  asymptotically periodic and develop a theory which relates these
  periodic growth coefficients to the waiting time distribution
  defining the periodic renewal process.  The first two coefficients,
  which are the mean frequency and effective diffusion coefficient of the number of
  events are considered in greater detail. They may be used to
  quantify stochastic synchronization.
\end{abstract}

\section{Introduction}
Many dynamical processes in physics, biology and chemistry, although
being inherently continuous, can be reduced to a series of discrete
events without losing much information.  Often this reduced
description in terms of discrete events has the additional property,
that time intervals between subsequent events are statistically
independent.  Such processes are called renewal processes \cite{cox}.
They are fully described by a waiting time distribution $w(\tau)$,
which governs the statistical properties of the time intervals $\tau$
between two subsequent events.  In the simplest case the probability
per unit time, i.e. the rate, for an event to occur is independent on
the time elapsed since the last event. This subclass of renewal
processes is called Markovian, because the event number of such a
process as a function of time is a Markovian stochastic process.  If
driven externally, the rate can be temporally modulated becoming thus
a function of the absolute time $t$, but still remains independent on
the waiting time $\tau$ elapsed since the previous event.

In the general case, the probability per unit time that an event will
happen is not independent of the time already elapsed since the last
event.  Then a non exponentially distributed waiting time between
subsequent events is observed which distinguishes these general
renewal processes from discrete Markovian processes.

The concepts of renewal processes have been successfully applied to
random sequences of spiking events in the dynamics of
neurons \cite{longtin,lindner,pakdaman},
to random walks in a
tilted periodic potential \cite{lindner2} and to the failure times of
biological and technical machineries \cite{cox2}.  In these models the
processes which generate the single events consist of a series of
several Markovian steps or possess a priori complicated waiting time
densities. For example, the generation of a new spike requires
excitation of the voltage variable over a threshold value which is
followed by the spiking and refractory time where neurons are unable
to produce new spikes. Also the ionic transport through pores involves
several steps as diffusional motion followed by escapes over potential
barriers. Molecular motors walk along microtubuli where sequences of
different configurational changes of the proteins realize one forward
step.

There is a long history of studies of stationary renewal processes
\cite{cox}, i.e. renewal processes having waiting time densities
$w(\tau)$, which do not depend on the absolute time $t$.  Also
stationary situations with multiple events and different stationary
waiting time statistics were studied.  Quantities of interest like the
mean event number in a certain period of time or the event number
diffusion are known in terms of the waiting time distribution, because
a fully developed theory exists for these time homogeneous processes
\cite{cox,cox2}

In connection with stochastic resonance (SR) the interest for
non stationary but periodic processes has grown \cite{SR,longtinSR,gammaitoni}.  In particular,
Markovian periodically driven models have been investigated
\cite{wiesenfeld}. The two state Markovian theory with
non stationary rates describes successfully the dynamic behavior in
bistable situations as shown in many different studies in various
fields of science. 
However, other systems like excitable dynamics as used for example to
describe the spiking mechanism in neuronal systems can no longer be
approximated by a simple discrete Markovian description.  If such
systems are periodically driven, as is for example the case in neurons
which respond to periodically varying inputs, more general concepts
are needed to describe their behavior.  In this paper we study the
general situation of non stationary but periodic renewal processes with
arbitrary waiting time distributions $w(\tau,t)$.  The periodicity is
reflected by a periodic dependence of the waiting time distribution on
the absolute time $t$ of the previous event.

There exist different possibilities to quantify the periodicity of a periodic
stochastic process, or, if one considers this periodicity as induced by
an external periodic signal, the quality of the response of the system
to this periodic signal. 
On the one hand, spectral based measures like the spectral power
amplification and the signal to noise ratio have been frequently
employed (see \cite{gammaitoni} and references therein). 
These spectral based response measures where also considered in the
context of periodically driven renewal
processes in \cite{goychuk,goychuk2} and for a special discrete state
model for excitable dynamics in \cite{prager2}.
To this end the sequence of events has to be somehow mapped
onto a stochastic process, e.g. by considering a sequence of delta
peaks located at the event times or by assigning alternatingly after
each event different values to the process.  
Another possibility to characterize periodicity of the process is to
consider the drift and diffusion properties  of the number of events. 
The evolution of the number of events in time can be characterized by two
quantities, the mean frequency of events and the effective diffusion
coefficient, which describes, how the variance of the number of events
grows in time and thus characterizes the regularity of the process.
The lower this effective diffusion coefficient for a fixed mean
frequency, the more regular, i.e. periodic, is the system's dynamics.
Both quantities together may therefore serve as a measure of
periodicity of the process and thus, if this periodicity stems form
the influence of a periodic input, as a measure of stochastic
synchronization between the driving signal and the system dynamics
\cite{janf,uspekhi,callenbach,morillo,prager}.

In this paper we present a method to calculate the mean frequency and
the effective diffusion coefficient of a periodically driven renewal
process in terms of its periodically time dependent waiting time
distribution $w(\tau,t)$.  After having derived the general concepts
and results in sections \ref{pdr} to \ref{driftdiff}, we consider two simple
situations, namely general but undriven renewal processes and
periodically modulated rate process, for which an explicit evaluation
of the general results is possible and agrees with the known results.
Finally in section \ref{fs} and \ref{toymod} we
numerically evaluate our theory for a toy model where the intervals between subsequent
events are governed by a a fixed but periodically varying waiting time
followed by a rate process with constant rate. The results agree with
simulations of the underlying periodic renewal process.  Many of the
calculations are summarized in a series of appendices.

\section{Periodically driven renewal processes}\label{pdr}
A periodically driven renewal process is a sequence of events occurring
at times $\ldots,t_i,t_{i+1},t_{i+2},\ldots$. The intervals between
two subsequent events are governed by the waiting time distribution
$w(\tau,t)$. 
The first argument
$\tau$ represents the waiting time whereas the second argument $t$
denotes the absolute time at which the last event happened. 
Therefore, $w(\tau,t_{i}) d\tau$ is the probability that the
event $i+1$ happens in the time interval $(t_i+\tau,t_i+\tau+d\tau]$
if event $i$ have been happened at time $t_i$. The dependence of the
waiting time distribution $w(\tau,t)$ on time $t$ is due to the
periodic driving and thus periodic with the period $\TT=2\pi/\Omega$
of the signal.  Although, in contrast to an ordinary renewal process,
the intervals between subsequent events are now correlated, it is
still only the time of the previous event which governs the statistics
of the following event, which justifies to stick to the term renewal.
Normalization holds at arbitrary time $t$ 
\begin{eqnarray*}
\int_0^{\infty}\,d\tau w(\tau,t) \,=\,1
\end{eqnarray*}

A lot of information about these processes is contained in the random
number of events $N_{t_0,t}$ which take place in the interval
$(t_0,t]$.  For example stochastic synchronization to the periodic
driving can be characterized as an integer relation between driving
frequency and the frequency of events and at the same time, a decrease
in the effective diffusion coefficient of the events, i.e. a more
regular (periodic) behavior.

We will evaluate the mean frequency and effective diffusion
coefficient based on the periodically time dependent waiting time
distribution $w(\tau,t)$, which defines the periodically driven
renewal process.  To this end we consider more generally the $n$th
cumulants $K^{(n)}_{t_0,t} $ of the event number $N_{t_0,t}$ and
define their increment per time
\begin{eqnarray}\label{defknt0}
  \kappa^{(n)}_{t_0,t}:=\lim_{\Delta t \to 0} \frac{K^{(n)}_{t_0,t+\Delta t}-K^{(n)}_{t_0,t}}{ \Delta t}
=\tde{t}K^{(n)}_{t_0,t}.
\end{eqnarray}
The first of these coefficients is the mean frequency while the second
is the effective diffusion coefficient. We show that asymptotically
\begin{eqnarray}\label{defkn1}
  \kappa^{(n)}(t):=\lim_{t_0 \to -\infty} \kappa^{(n)}_{t_0,t}
\end{eqnarray}
become periodic functions of time with the period of the external
driving.  

To this end we introduce following \cite{stratonovich,vankampen} the generating
functional $L_{t_0,t}[v]$ of the considered driven renewal process
as
\begin{eqnarray*}
  L_{t_0,t}[v]=\mean{\prod_{i=1}^{N_{t_0,t}} (1+v(t_i))}
\end{eqnarray*}
where $t_i$ are the  times of the 
events in the interval $(t_0,t]$.  Then the $n$th moment $
M^{(n)}_{t_0,t}$ of event number $N_{t_0,t}$ is given by
\begin{eqnarray}\label{meanNk}
  M^{(n)}_{t_0,t}:=\mean{N_{t_0,t}^n}=\pd{n}{u}L_{t_0,t}[e^{u}-1]\Big|_{u=0}.
\end{eqnarray}
The generating functional $L_{t_0,t}$ can be expressed in terms of
the distribution functions $f_s(t_1,\ldots,t_s)$, which 
govern the probability
\begin{eqnarray*}
  dP=f_s(t_1,\ldots,t_s)dt_1 \ldots dt_s
\end{eqnarray*}
to find one event in each of the intervals $(t_i,t_i+d
t_i),\;i=1,\ldots,s$ regardless of how many events are outside these
intervals,
as \cite{stratonovich,vankampen}
\begin{eqnarray}\label{Lf}
L_{t_0,t}[v]=1+\sum_{s=1}^\infty 
\int_{t_0}^{t} \!\!\!d\tau_1\int_{t_0}^{\tau_1} \!\!\!d\tau_2\ldots \int_{t_0}^{\tau_{s-1}}
  \!\!\!d\tau_s \, f_s(\tau_1,\tau_2,\ldots,\tau_s)\,   v(\tau_1)\ldots v(\tau_s)\;.\nonumber\\
\end{eqnarray}
The generating functional can also be expressed in terms of the
correlation functions $g_s(t_1,\ldots,t_s)$ as
\begin{eqnarray}\label{Lg}
\lefteqn{  L_{t_0,t}[v]=\exp\Big[}\nonumber\\&&\sum_{s=1}^\infty \int_{t_0}^{t}
  d\tau_1\int_{t_0}^{\tau_1} d\tau_2\ldots \int_{t_0}^{\tau_{s-1}}
  d\tau_s \, g_s(\tau_1,\tau_2,\ldots,\tau_s)\,   v(\tau_1)\ldots v(\tau_s) \Big]\,.\;\;
\end{eqnarray}
Eq. (\ref{Lg}) together with eq. (\ref{Lf}) define 
 the correlation functions in terms of the distribution functions.

According to eq. (\ref{Lg}) the moments eq. (\ref{meanNk}) can be
expressed as
\begin{eqnarray}
\label{Moments}
  M^{(n)}_{t_0,t}&=&\pd{n}{u}
\exp\Big[\sum_{s=1}^\infty G_s(t_0,t) (e^u-1)^s \Big] \Big|_{u=0}
\end{eqnarray}
where
\begin{eqnarray*}
  G_s(t_0,t):=\int_{t_0}^{t} d\tau_1\int_{t_0}^{\tau_1} d\tau_2\ldots \int_{t_0}^{\tau_{s-1}}
  d\tau_s \, g_s(\tau_1,\tau_2,\ldots,\tau_s)\, .
\end{eqnarray*}
From formula (\ref{Moments}) we can evaluate the corresponding
cumulants $K^{(n)}_{t_1,t_2}$ as (see appendix \ref{relmomcum})
\begin{eqnarray}\label{cumGs}
  K^{(n)}_{t_0,t}&=&\pd{n}{u}\sum_{s=1}^\infty 
G_s(t_0,t) (e^u-1)^s \Big|_{u=0}.
\end{eqnarray}

As the considered renewal processes are periodic in time with period
$\TT$, the distribution functions and therefore also the correlation
functions are likewise periodic in time,
\begin{eqnarray}\label{periodicitygs}
  g_s(\tau_1,\ldots,\tau_s)=g_s(\tau_1+\TT,\ldots,\tau_s+\TT).
\end{eqnarray}
Then the time derivative of the function $G_s(t_0,t)$ yields
\begin{eqnarray*}
  \tde{t}G_s(t_0,t)&=& \int_{t_0}^{t} d\tau_2\ldots \int_{t_0}^{\tau_{s-1}}  d\tau_s g_s(t,\tau_2,\ldots,\tau_s) \\
&=& \frac{1}{(s-1)!} \int_{t_0}^{t} d\tau_2\ldots \int_{t_0}^{t}
  d\tau_s g_s(t,\tau_2,\ldots,\tau_s)\\
&=&\frac{1}{(s-1)!} \int_{0}^{t-t_0} d\tau_2\ldots \int_{0}^{t-t_0}
  d\tau_s g_s(t,t-\tau_2,\ldots,t-\tau_s)
\end{eqnarray*}
which can be expressed in the asymptotic limit as
\begin{eqnarray}\label{periodicGs}
  \lim_{t_0\to-\infty}\tde{t} G_s(t_0,t) =  \frac{1}{(s-1)!}\int_{0}^{\infty} d\tau_2\ldots \int_{0}^{\infty}
  d\tau_s g_s(t,t-\tau_2,\ldots,t-\tau_s)\,.
\end{eqnarray}
To ensure that this limit exists, we additionally suppose that
$g_s(\tau_1,\ldots,\tau_s)$ decreases sufficiently fast to zero
for any pair of time difference $|\tau_i-\tau_j|\to \infty$. In
\cite{vankampen} this property is called \emph{cluster property},
while for stationary systems this property is called ergodicity.

According to eq. (\ref{periodicitygs}) the asymptotic time derivative
eq. (\ref{periodicGs})
is a periodic function in $t$ and thus (cf. eq. (\ref{cumGs})) the
coefficients
\begin{eqnarray}\label{defkn}
  \kappa^{(n)}(t)=\lim_{t_0\to -\infty }\tde{t} K^{(n)}_{t_0,t}.
\end{eqnarray}
are periodic in time, as well.

The time dependent waiting time distribution can be expressed in terms
of the distribution functions as \cite{vankampen}
\begin{eqnarray*}
  w(\tau,t)=f_1(t+\tau)+\sum_{s=1}^\infty \frac{(-1)^s}{s!}
  \int_t^{t+\tau}dt_1\ldots dt_s \;f_{s+1}(t+\tau,t_1,\ldots,t_s).
\end{eqnarray*}
However, we are faced with the inverse problem, namely to express  the
distribution functions $f_s$ or likewise the correlation functions
$g_s$ in terms of the waiting time distribution $w(\tau,t)$ in
order to finally evaluate the coefficients $\kappa^{(n)}(t)$ according to
eqs. (\ref{defkn}) and (\ref{cumGs}). Such a relation is not
known to the authors and even if it exists, the explicit evaluation of
the infinite sum in eq. (\ref{cumGs}) will be challenging.
Thus to obtain the $\kappa^{(n)}(t)$ in terms of the time dependent waiting
time distribution $w(\tau,t)$ we have to adopt a different
approach. 


\section{The microscopic  master equation}
Let us consider the probabilities $p_k(t)$ to have had $k$ events up
to time $t$. Furthermore let $j_k(t)$ be the probability flux from
state $k$ to state $k+1$, i.e. the probability per time that the
$k+1^\text{st}$ event happens at time $t$. This probability obeys the
continuity equation
\begin{equation}
  \label{eq:flux}
  \tde{t} p_k(t)\,=\,j_{k-1}(t)\,-\,j_k(t).
\end{equation}
If we further assume as initial condition that event 1 happened at
time $t_0$, i.e. $j_0(t)=\delta(t-t_0)$,  the relation between the
probability fluxes of the renewal process can be expressed by the ``microscopic'' dynamics
as (for the undriven case see e.g. \cite{gillespie})
\begin{equation}
\label{dyneq}
   j_k(t)\,=\,\int_{t_0}^t dt' j_{k-1}(t') w(t-t',t'),\quad k\ge 1.  
\end{equation}
Using this relation  one readily obtains from the continuity equation (\ref{eq:flux})  
\begin{equation}\label{relprobflux}
  p_k(t)\,=\,\int_{t_0}^t dt' j_{k-1}(t') z(t-t',t'),\quad k\ge 1
\end{equation}
where $z(\tau,t)=1-\int_0^\tau d\tau' w(\tau',t)$ is the probability
to wait longer than $\tau$ until the next event, if the last event
happened at $t$.  
In case of a Markovian
renewal process with time dependent rate $\gamma(t)$ the probability
flux $j_k$ is related to the probability $p_k$ by $j_k(t)=\gamma(t)
p_k(t)$. Thus in this case the dynamics can be completely expressed in
terms of the probabilities $p_k$. In the general case however we need
a formulation in terms of $p_k$ and $j_k$ as expressed in eqs.
(\ref{relprobflux}) and (\ref{dyneq}).  

The moments of the number of
events $\tilde N_{t_0,t}$ in the interval $(t_0,t]$ can be expressed
in terms of the $p_k$ as
\begin{eqnarray}\label{tildemom}
  \tilde M^{(n)}_{t_0,t}=\sum_{k=0}^\infty k^n p_k(t).
\end{eqnarray}
Note that the moments $\tilde M^{(n)}_{t_0,t}$ differ from the moments
defined by eq. (\ref{meanNk}) since the corresponding event number
$\tilde N_{t_0,t}$ is conditioned on having had an event at time $t_0$
in contrast to $N_{t_0,t}$. However the asymptotic behavior of both
families of moments agrees.  

In principle one can calculate the cumulants from the moments eq.
(\ref{tildemom}) to obtain eventually the coefficients
$\kappa^{(n)}(t)$ according to eq. (\ref{defkn}).  However, in
practice this is not feasible in general, as one has to calculate an
infinite sum over the $p_k(t)$ where each $p_k$, according to eqs.
(\ref{relprobflux}) and (\ref{dyneq}), is a $k$-fold integral
involving the waiting time distributions $w(\tau,t)$ and the
corresponding survival probabilities $z(\tau,t)$.

\section{Embedding the discrete dynamics into a continuous one}

\begin{figure}[htbp]
  \centering\includegraphics[width=9cm]{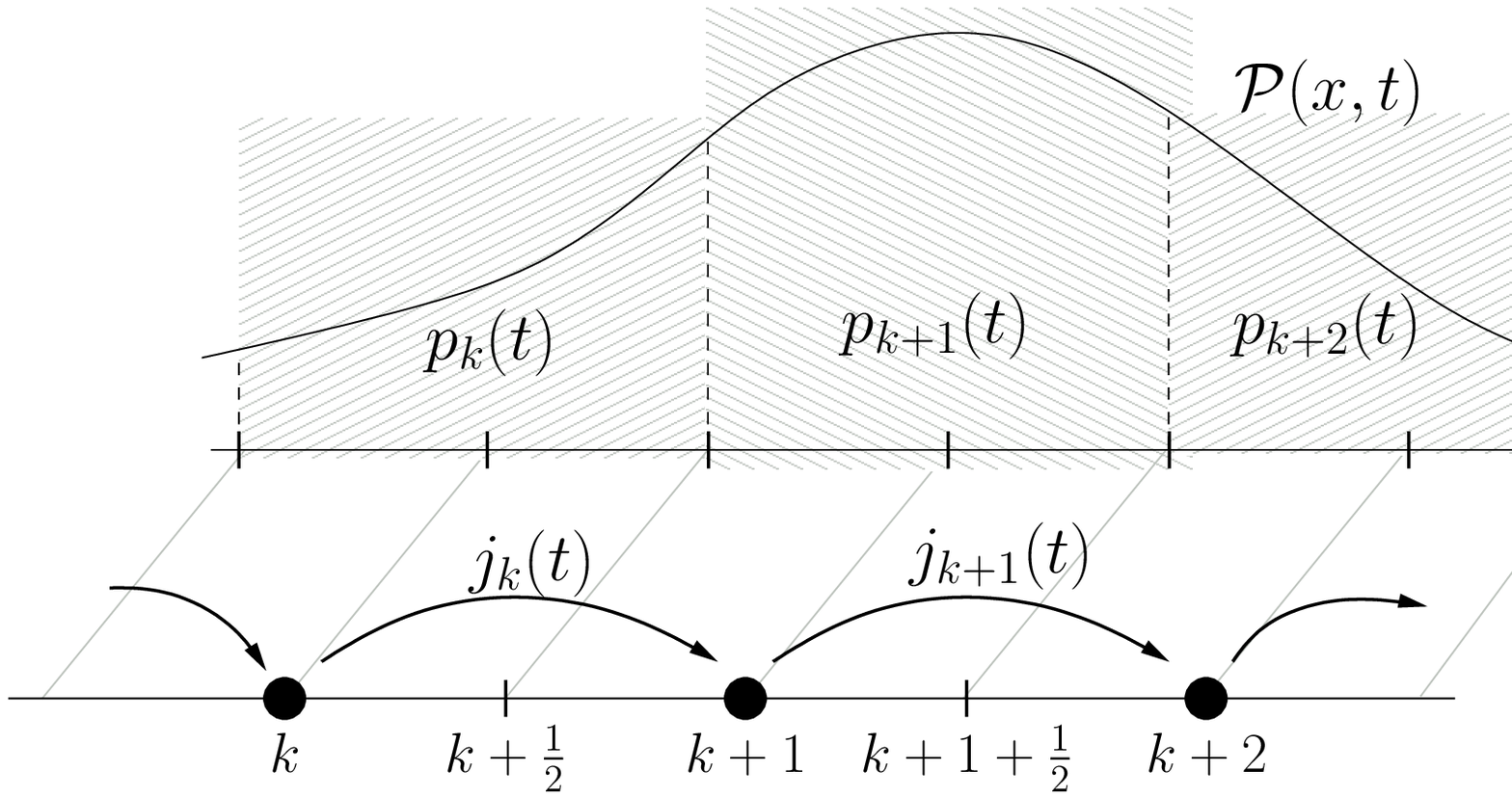}
\caption{\label{sketch}
Schematic view of the discrete event dynamics $p_k(t)$ and the continuous
description in terms of $\contvar$.}
\end{figure}

To find a simpler relation between the periodic coefficients
$\kappa^{(n)}(t)$ and the time dependent waiting time $w(\tau,t)$,
which governs the microscopic dynamics, we construct a continuous
embedding in the asymptotic limit $t_0\to -\infty$. Consider the
envelope $\PP(\contvar,t)$ of the discrete probabilities $p_k(t)$ as a
probability distribution on a continuous state space.  Thus, respecting
the normalization we adopt the relation (cf.  Fig. \ref{sketch})
\begin{eqnarray}\label{ansatzp}
p_k(t)=\int_{k-\frac{1}{2}}^{k+\frac{1}{2}}d \contvar\PP(\contvar,t).
\end{eqnarray}
Assuming that the variation of
$\PP(x,t)$ within an interval $(k-\frac{1}{2},k+\frac{1}{2})$ becomes
arbitrary small for large $t$ the difference between the moments and therefore also the
cumulants of both the discrete and the continuous process tends to zero in this
asymptotic limit. 
Accordingly, also the change in time of the cumulants of the continuous
process have to agree with the corresponding quantities
$\kappa^{(n)}(t)$ of the discrete process.

One possible choice  to formulate a dynamics of the continuous envelope
$\PP(\contvar,t)$ is  a Kramers-Moyal equation.
This choice renders  the underlying stochastic
process $\contvar(t)$  Markovian.  To ensure the equality of the increments of
the cumulants, the coefficients of the Kramers-Moyal expansion have to
coincide with the coefficients $\kappa^{(n)}(t)$ of the renewal process (cf.
appendix \ref{growthofcummulants}, \cite{stratonovich}), i.e.
\begin{eqnarray}\label{generalfpe}
  \pde{t} \PP(\contvar,t)=\sum_{n=1}^\infty \frac{(-1)^n}{n!}
  \kappa^{(n)}(t) \pd{n}{\contvar}  \PP(\contvar,t) 
\end{eqnarray}

The probability current $j_k(t)$ in the discrete system is
correspondingly related to the probability current of the continuous
envelope description $\JJ(\contvar,t)$. 
According to the relation between the discrete and continuous
probability eq. (\ref{ansatzp}),  the discrete probability current $j_k(t)$ from $k$ to $k+1$ is
equal to the continuous probability current $\JJ(\contvar,t)$ at $\contvar=k+\frac{1}{2}$,
\begin{eqnarray}\label{ansatzj}
  j_k(t)=\JJ(k+\frac{1}{2},t).
\end{eqnarray}


The continuous probability current $\JJ(\contvar,t)$ is related to the probability
distribution $\PP(\contvar,t)$ by the continuity equation
\begin{eqnarray}\label{probcurrent}
  \pde{t}\PP(\contvar,t)&=&-\pde{\contvar} \JJ(\contvar,t).
\end{eqnarray}
and therefore according to eq. (\ref{generalfpe})
\begin{eqnarray}\label{probcurrent2}
  \JJ(\contvar,t)=-\sum_{n=1}^\infty \frac{(-1)^n}{n!} \kappa^{(n)}(t) \pd{n-1}{\contvar}  \PP(\contvar,t)\,.
\end{eqnarray}
Thus from eq. (\ref{ansatzj}) we deduce
\begin{eqnarray}\label{ansatzj2}
  j_k(t)=-
\sum_{n=1}^\infty \frac{(-1)^n}{n!} \kappa^{(n)}(t) \pd{n-1}{\contvar}
\PP(\contvar ,t)\Big|_{\contvar=k+\frac{1}{2}}
\end{eqnarray}

Finally we want to mention that the more general embedding of the
discrete process
\begin{eqnarray}\label{ansatzpj2}
p_k(t)=\int_{k-1+l}^{k+l}d \contvar\PP(\contvar,t)\qquad
\text{and}\qquad j_k(t)=\JJ(k+l,t),\quad l \text{ arbitrary}.
\end{eqnarray}
leads to the same results as the embedding chosen in
eqs. (\ref{ansatzp}) and (\ref{ansatzj}), which corresponds to
choosing $l=\frac{1}{2}$ in eq. (\ref{ansatzpj2}).
 

\section{The asymptotic drift and the diffusion coefficient}\label{driftdiff}
Having fixed the relation between the probabilities and probability fluxes
of the discrete renewal process and the continuous embedding,
it is now possible to relate the coefficients $\kappa^{(n)}(t)$
appearing in the continuous description (\ref{generalfpe}) to the waiting time distribution
$w(\tau,t)$  of the renewal process, involved in the microscopic dynamics
(\ref{dyneq}) and (\ref{relprobflux}). 
As we are considering the asymptotic behavior we have to  pass to the
asymptotic limit in eqs.  (\ref{relprobflux}) and (\ref{dyneq}) by
shifting the initial time  $t_0\to -\infty$. This results in
\begin{eqnarray}\label{relprobfluxasymp}
  p_k(t)=\int_0^\infty d\tau j_{k-1}(t-\tau) z(\tau,t-\tau)
\end{eqnarray}
and
\begin{eqnarray}\label{dyneqasymp}
  j_k(t)= \int_0^\infty d\tau j_{k-1}(t-\tau) w(\tau,t-\tau).
\end{eqnarray}
Inserting eqs. (\ref{ansatzp}) and (\ref{ansatzj2}) into the above eq.
(\ref{relprobfluxasymp}) we end up with
\begin{eqnarray}\label{interstep}
\int_{-\frac{1}{2}}^{\frac{1}{2}}d \Delta \contvar
\PP(\contvar-\Delta \contvar,t)&=&-\int_0^\infty d\tau z(\tau,t-\tau)\\&&
 \sum_{n=1}^\infty \frac{(-1)^n}{n!} \kappa^{(n)}(t-\tau)
 \pd{n-1}{\contvar} \PP(\contvar-\frac{1}{2},t-\tau)\nonumber
\end{eqnarray}
with  $\contvar=k$.
The probability  $\PP(\contvar-\Delta \contvar,t-\tau)$ can be
expressed in terms of the probability $\PP(x,t)$ and its derivatives 
$\pd{m}{\contvar} \PP(\contvar,t)$ by
performing a Taylor expansion of $\PP(\contvar-\Delta\contvar,t-\tau)$
around $x,t$ and converting the time derivatives to derivatives with
respect to the state $\contvar$ using the Kramers-Moyal equation (\ref{generalfpe}). 
This results in (cf. appendix \ref{herleitung})
\begin{eqnarray*}
\PP(\contvar-\Delta\contvar,t-\tau)&=&\PP(\contvar,t) +
c^{(1)}_t(t-\tau,\Delta\contvar)\pde{\contvar}\PP(\contvar,t)+ O(2)
\end{eqnarray*}
with 
\begin{eqnarray*}
c_t^{(1)}(t-\tau,\Delta\contvar)&=&\int_{0}^\tau d\tau' \kappa^{(1)}(t-\tau')-\Delta \contvar \,.
\end{eqnarray*}
Above $O(2)$ denotes terms proportional to
$\pd{m}{\contvar}\PP(\contvar,t)$, $m\ge 2$. 

Denoting by $L$ the characteristic length scale over which the probability
distribution $\PP(\contvar,t)$ varies, we may introduce the new variable $\tilde
\contvar =\frac{\contvar}{L}$. Then
\begin{eqnarray*}
  \pd{m}{\contvar}\PP(\contvar,t)=
  \frac{1}{L^m}\pd{m}{\tilde\contvar}\PP(\tilde \contvar,t)\Big|_{\tilde
  \contvar=\frac{\contvar}{L}}
\end{eqnarray*}
Due to the dispersion of the probability distribution
$\PP(\contvar,t)$ 
the characteristic length $L$ asymptotically goes to infinity.

Equating the coefficients
of $\PP(\contvar,t)$ and
$\pde{\contvar}\PP(\contvar,t)$ 
on both sides of eq.  (\ref{interstep}), or equivalently, considering 
the 0th and 1st order in $L$,  we end up with
\begin{eqnarray}
 \int_0^\infty d\tau \kappa^{(1)}(t-\tau)z(\tau,t-\tau)&=&1\label{omega}\\
  \frac{1}{2}\int_0^\infty d\tau
  \kappa^{(2)}(t-\tau)z(\tau,t-\tau)&=&\nonumber\\&&\hspace{-3cm}\int_0^\infty d\tau
  \kappa^{(1)}(t-\tau)\big[\int_{0}^\tau d\tau'\kappa^{(1)}(t-\tau')
-\frac{1}{2}\big]z(\tau,t-\tau)\,.\qquad\label{D}
\end{eqnarray}
Eq. (\ref{D}) can be further simplified using eq. (\ref{omega}),
which leads to
\begin{eqnarray}\label{D2}
\int_0^\infty d\tau \kappa^{(2)}(t-\tau)z(\tau,t-\tau)&=&\nonumber\\
&&\hspace{-3cm}2 \int_0^\infty d\tau
  \kappa^{(1)}(t-\tau)\int_{0}^\tau
  d\tau'\kappa^{(1)}(t-\tau')z(\tau,t-\tau)-1\,.\quad
\end{eqnarray}
These two expressions, relating the asymptotic drift and diffusion
properties  of a periodically driven
renewal process as
expressed by $\kappa^{(1)}(t)$ and $\kappa^{(2)}(t)$  
to its microscopic properties defined by $z(\tau,t)$
present the cornerstone result of our paper.  Equations which govern
the higher cumulant growth coefficients $\kappa^{(n)}(t),\;n\ge 3$ can also
be derived using this method by evaluating the coefficients of 
higher order derivatives of $\PP(\contvar,t)$ (for $\kappa^{(3)}(t)$
see appendix \ref{K3}).

In a former work \cite{prager} we investigated stochastic
synchronization in a non Markovian two state model for excitable
systems. The result presented in this paper for the mean velocity and
effective diffusion coefficient can be shown to be equivalent to the
results presented here (see appendix \ref{equivts}).

Often one is not directly interested in the number of events but in
some quantity proportional to the number of events, like a phase,
which increases by $\intrinsicl=2 \pi$ for each event, or a position
which increases by some length $\intrinsicl=l$ if we consider a
unidirectional random walk with step-length $l$ whose steps are
governed by the described renewal dynamics. In these cases the $i$th
cumulant growth coefficient $\kappa^{(i)}(t)$ is scaled by
$\intrinsicl^i$.  For $\kappa^{(1)}(t)$ and  $\kappa^{(2)}(t)$  
this is achieved multiplying the constant term 1 on the right hand side of eqs. (\ref{omega}),
and  (\ref{D2})  by $\intrinsicl$ or  $\intrinsicl^2$ respectively.

Finally one may ask, why it is justified to prescribe a continuous
Markovian envelope dynamics to an inherently non Markovian discrete
process.  The obvious idea, that the non Markovian nature of the discrete
process is rendered Markovian by being mapped onto an extended
continuous state space is misleading. The point is, that the continuous
Markovian process $\contvar(t)$ as described by eq. (\ref{generalfpe})
is not an envelope dynamics of  the full discrete non Markovian process,
but it only covers the asymptotic behavior of the non Markovian
process, i.e the regime, where the discrete
process has forgotten its initial preparation.  

\section{Comparison with known results for undriven renewal processes
  and periodically driven rate processes}\label{sec:comp}
Let us evaluate expressions (\ref{omega}) and (\ref{D}) for an
undriven renewal process, i.e. $z(\tau,t)\equiv z(\tau)$. Then it
follows that $\kappa^{(1)}(t)$ is constant and eq. (\ref{omega}) leads to
\begin{eqnarray*}
  \kappa^{(1)}(t)=\frac{\intrinsicl}{\mean{T}}
\end{eqnarray*}
with 
\begin{eqnarray*}
\mean{T^n}:=\int_0^\infty d\tau \tau^n w(\tau).
\end{eqnarray*}
To derive this result we have used the fact that 
\begin{eqnarray*}
  \int_0^\infty d\tau \tau^n z(\tau)
&=&\int_0^\infty d\tau
  \frac{\tau^{n+1}}{n+1}w(\tau)=\frac{1}{n+1}\mean{T^{n+1}}
\end{eqnarray*}
which holds if $z(\tau)$ decreases sufficiently fast  for $\tau\to\infty$.
Accordingly eq. (\ref{D2}) gives
\begin{eqnarray*}
  \kappa^{(2)}(t)&=&\intrinsicl^2\frac{\mean{T^2}-\mean{T}^2}{\mean{T}^3}
\end{eqnarray*}
which agrees with the known results for stationary renewal processes \cite{cox}. 
The corresponding expression for $\kappa^{(3)}(t)$ is presented in
appendix \ref{K3}.

Next we consider a periodically driven Markov process, i.e.
 \begin{eqnarray*}
   w(\tau,t)=\gamma(t+\tau)\exp\big(-\int_t^{t+\tau} d\tau' \gamma(\tau')\big)
 \end{eqnarray*}
and
\begin{eqnarray*}
  z(\tau,t)=\exp\big(-\int_t^{t+\tau} d\tau' \gamma(\tau')\big).
\end{eqnarray*}
Then it can be easily shown that eq. (\ref{omega}) is solved by
\begin{eqnarray*}
  \kappa^{(1)}(t)=\intrinsicl \gamma(t).
\end{eqnarray*}
The first term on the right hand side og eq. (\ref{D2}) can also be
simplified in this case using  integration by parts to give
$\intrinsicl^2$.
Therefore $\kappa^{(2)}(t)$ is governed by
\begin{eqnarray*}
  \int_0^\infty d\tau \kappa^{(2)}(t-\tau)z(\tau,t-\tau)&=&\intrinsicl^2
\end{eqnarray*}
which is solved by
\begin{eqnarray*}
  \kappa^{(2)}(t)=\intrinsicl^2 \gamma(t)
\end{eqnarray*}

For more complicated processes with general time dependent waiting
time distributions eqs. (\ref{omega}) and (\ref{D2}) can only be solved
numerically for the periodic solution.

\section{Numerical solution in Fourier space}\label{fs}
As eqs. (\ref{omega}) and (\ref{D}) cannot be solved analytically for
arbitrary waiting time distributions $w(\tau,t-\tau)$ and
corresponding survival probabilities $z(\tau,t-\tau)$ one has to
resort to numerical methods.  To this end we perform a Fourier
expansion of the periodic function $\kappa^{(1)}(t)$ and analogously for
$\kappa^{(2)}(t)$,
\begin{eqnarray}\label{fourierdecompkappa1}
   \kappa^{(1)}(t)=\sum_{k=-\infty}^\infty \kappa^{(1)}_k \exp(i k \Omega t),\qquad
   \kappa^{(1)}_k=\frac{1}{\TT} \int_0^{\TT} dt \kappa^{(1)}(t) \exp(-i k \Omega t),
\end{eqnarray}
where $\Omega=2\pi/\TT$ is the frequency of the external driving.  We
further expand the survival probability $z(\tau,t)$ with respect to
the second periodic argument as
\begin{eqnarray*}
  z(\tau,t)=\sum_{k=-\infty}^\infty z_k(\tau) \exp(ik\Omega t)\qquad
   z_k(\tau)=\frac{1}{\TT} \int_0^{\TT} d t z(\tau,t) \exp(-i k \Omega t)
\end{eqnarray*}
Abbreviating 
\begin{eqnarray}\label{fouriercoeffs}
  z_{k,l}=\int_0^\infty d\tau z_k(\tau) \exp(-il \Omega \tau)\;,\qquad
  h_{k,l}=\int_0^\infty d\tau \tau z_k(\tau) \exp(-il \Omega \tau)
\end{eqnarray}
eq. (\ref{omega}) can be written as
\begin{eqnarray}\label{omegafourier}
  \sum_{k=-\infty}^\infty \kappa^{(1)}_{k} z_{m-k,m}=\intrinsicl \delta_{m,0}, \quad
  m=-\infty,\ldots,\infty
\end{eqnarray}
while eq. (\ref{D2}) reads
\begin{eqnarray}\label{Dfourier}
\lefteqn{  \sum_{k=-\infty}^\infty \kappa^{(2)}_{k} z_{m-k,m}=
2 \sum_{k=-\infty}^\infty }\\&&\Big[\big[\sum_{l=-\infty,l\neq 0}^\infty
\frac{\kappa^{(1)}_l \kappa^{(1)}_k}{il\Omega}(z_{m-k-l,m-l}-z_{m-k-l,m})\big]+\kappa^{(1)}_0
\kappa^{(1)}_k  h_{m-k,m}\Big]\nonumber\\&&-\intrinsicl^2 \delta_{m,0}\nonumber.
\end{eqnarray}
The corresponding equation for the third coefficient $\kappa^{(3)}(t)$
in Fourier space is presented in appendix \ref{K3}.

These infinite dimensional inhomogeneous linear equations can then be
numerically solved, after being truncated to a finite dimensional
system.

\section{A simple example--comparison between theory and simulations}\label{toymod}
\begin{figure}[htbp]
  \centering\includegraphics[width=9cm]{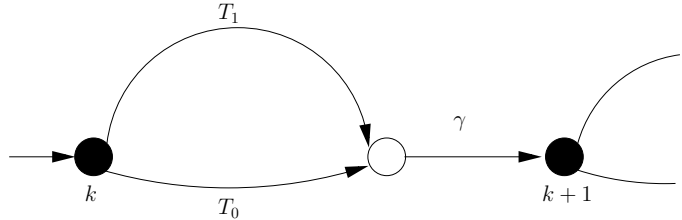}
\caption{\label{toymodelsketch}
Depending on whether the periodic signal is in the first or second half
period the system either waits the fixed time $T_0$ or $T_1$. In both
cases the system waits an additional exponentially with rate $\gamma$ 
distributed time.}
\end{figure}

\begin{figure}[htbp]
  \centering\includegraphics[width=9cm]{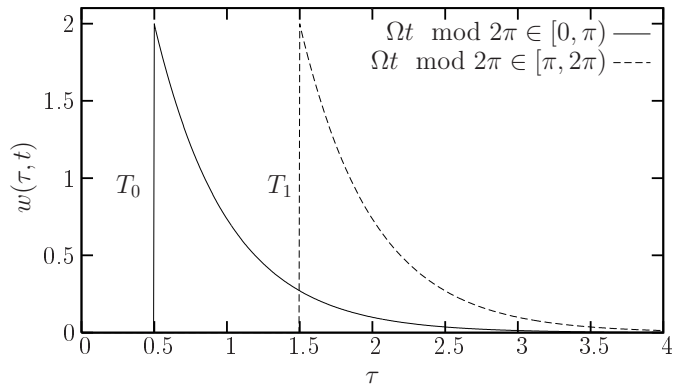}
\caption{\label{wtdtoymodel}
Waiting time distribution of the toy model eq. (\ref{toymodel}) with
dichotomic fixed waiting time from eq. (\ref{toymodelT}) for
$T_0=0.5$, $T_1=1.5$ and $\gamma=2$.
Depending on the signal phase of the event, the system waits either a
long (dashed line) or a short time(solid line) plus an exponentially
distributed time until the next event.
}
\end{figure}

Consider the toy renewal process, where the time between subsequent
events is composed of a fixed waiting time, which depends on the
signal phase of the previous event and a rate process with rate
$\gamma$. The waiting time distribution is thus given by
\begin{eqnarray}\label{toymodel}
\nonumber
  w(\tau,t)=\theta(\tau-T(t)) \gamma e^{-\gamma(\tau-T(t))}.
\end{eqnarray}
Suppose further that the fixed waiting time is either $T_0$
or $T_1$ depending on whether the signal phase of the previous event
was within $[0,\pi)$ or $[\pi,2\pi)$, i..e
\begin{eqnarray}\label{toymodelT}
\nonumber
 T(t)=\left\{
    \begin{array}{ll}
      T_0&\text{ if }
      \Omega t\mod 2\pi \in [0,\pi)\\
      T_1&\text{ if } \Omega t\mod 2\pi \in [\pi,2\pi)
    \end{array}
\right.\\&&
\end{eqnarray}
A sketch of this system is shown in Fig. \ref{toymodelsketch} while
 corresponding waiting time distribution is plotted in Fig. \ref{wtdtoymodel}

\begin{figure}[htbp]
  \centering
 \includegraphics[width=10cm]{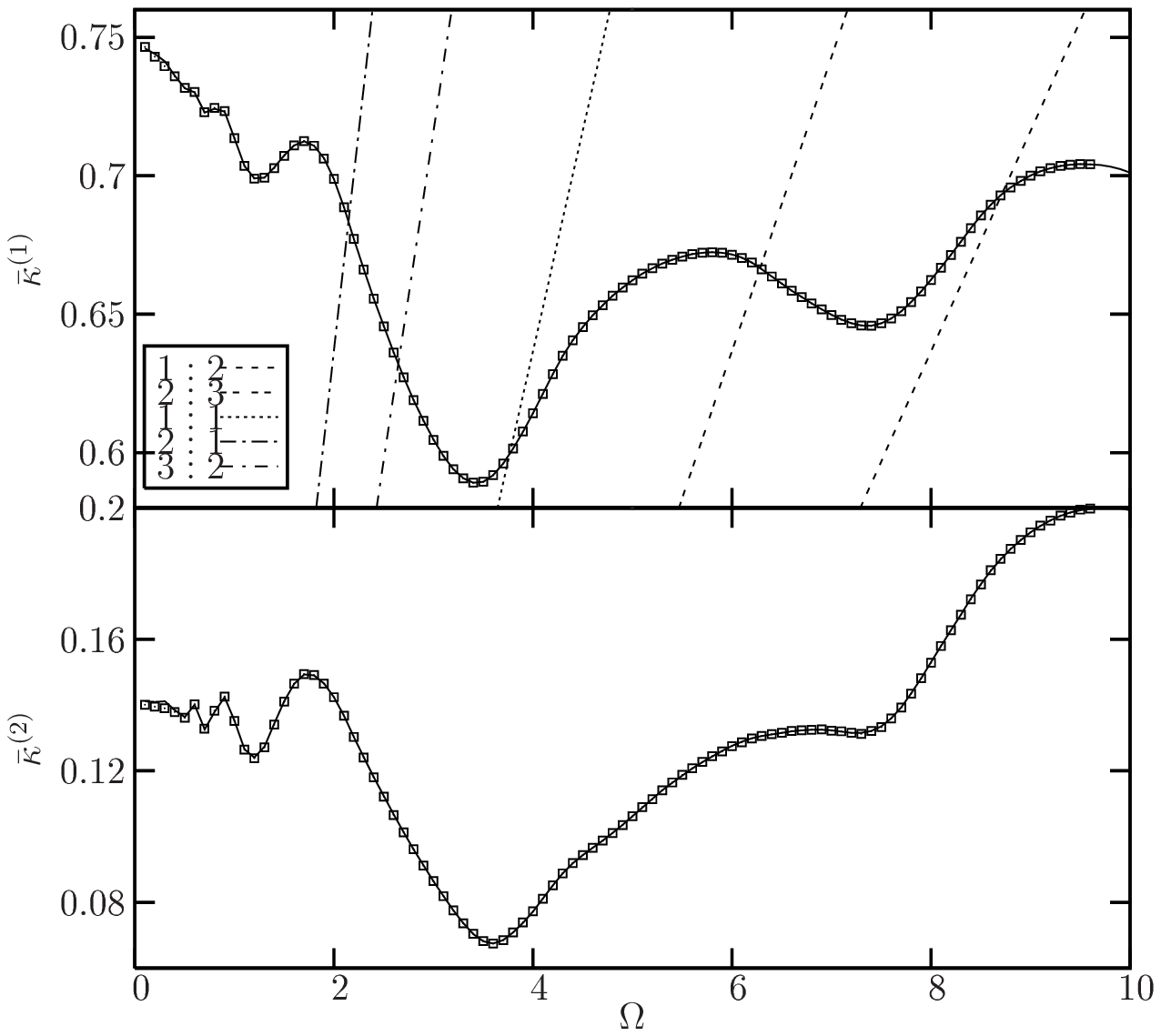} 
\caption{  
\label{comptheosim}
Comparison of the period averaged mean frequency
$\bar \kappa^{(1)}=\frac{1}{\TT}\int_0^{\TT} dt \kappa^{(1)}(t)$ and  the period averaged
effective diffusion coefficients $\bar \kappa^{(2)}=\frac{1}{\TT}\int_0^{\TT} dt
\kappa^{(2)}(t)$  where $\TT=2\pi/\Omega$ is the period of the signal,
 for the toy model eq. (\ref{toymodel}) with $T_0=0.5$, $T_1=1.5$ and
 $\gamma=2$.
The solid lines are results of the theory eqs. (\ref{omega}) and
(\ref{D2}), numerically evaluated according to
eqs. (\ref{omegafourier}) and (\ref{Dfourier}) truncated to 40 coefficients,
while the symbols are obtained from direct simulations of
the driven renewal process. The straight lines in the upper plot
indicate $n:m$ relations between system frequency and signal
frequency, i.e. frequency locking. Clearly the system does not
show this behavior.
}
\end{figure}

\begin{figure}[htbp]
  \centering
 \includegraphics[width=10cm]{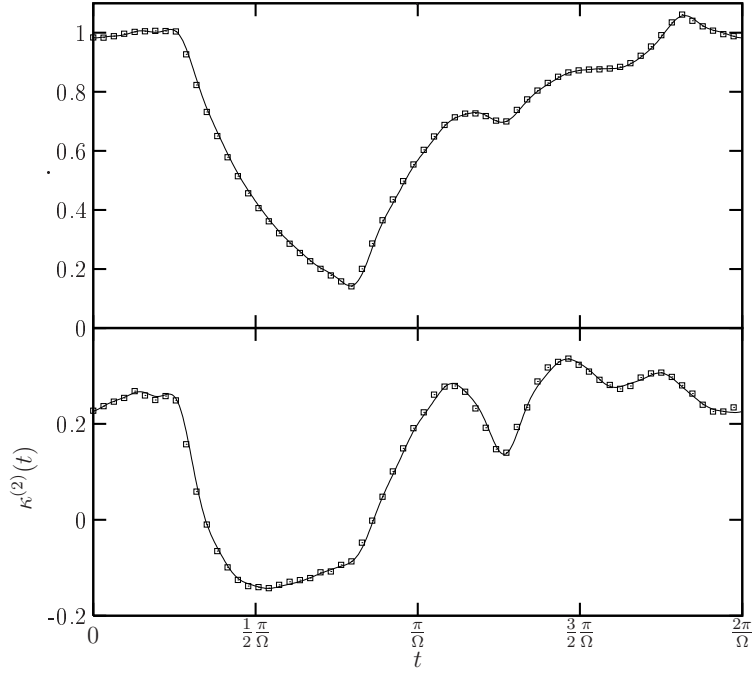} 
\caption{  
\label{comptheosimt}
Comparison of the  mean frequency $\kappa^{(1)}(t)$ and  the 
effective diffusion coefficients $\kappa^{(2)}(t)$  for
 the toy model eq. (\ref{toymodel}) with $T_0=0.5$, $T_1=1.5$ and
 $\gamma=2$ and $\Omega=1.7$.
The solid lines are results of the theory eqs. (\ref{omega}) and
(\ref{D2}), numerically evaluated according to
eqs. (\ref{omegafourier}) and (\ref{Dfourier}) truncated to 20 coefficients,
while the symbols are obtained from direct simulations of
the driven renewal process. 
}
\end{figure}

The corresponding Fourier coefficients $z_{k,l}$ and $h_{k,l}$ as
presented in eqs.
(\ref{fouriercoeffs}) can be analytically evaluated for this waiting
time distribution, however the final results, being too long and at the
same time yielding not much information, will not be presented
here.  Having evaluated these Fourier coefficients, we calculated the
mean frequency and the effective diffusion coefficient according
to eqs. (\ref{omegafourier}) and (\ref{Dfourier}) using LAPACK to
solve these linear equations.  
The results are compared to simulations
of the renewal process in Figs.\ref{comptheosim} and
\ref{comptheosimt}, showing perfect agreement.

The  mean velocity $\bar \kappa^{(1)}=\frac{1}{\TT}\int_0^{\TT} dt
\kappa^{(1)}(t)$ and effective diffusion coefficient
$\bar \kappa^{(2)}=\frac{1}{\TT}\int_0^{\TT} dt
\kappa^{(2)}(t)$ 
can be used to characterize stochastic synchronization
of the process to the periodic signal.
These synchronization regions are defined  by  a rational relation between system frequency and signal
frequency and a minimum of the effective diffusion coefficient.
Although this model shows minima of the effective diffusion
coefficient as a function of the driving frequency Fig. \ref{comptheosim}, these minima do
not correspond to frequency synchronization as the system frequency and
the signal frequency are not proportional. This stands in contrast to a
similar model with a fixed waiting time and a dichotomically
periodically modulated rate, used to describe periodically driven
excitable systems \cite{prager}. In this model we found several
different $n:m$ synchronization regions.
Also the full periodically time dependent coefficients $\kappa^{(1)}(t)$ and
$\kappa^{(2)}(t)$ as determined by our theory (\ref{omega}) and (\ref{D2})
agree with results taken from simulations of the underlying renewal
process Fig. \ref{comptheosimt}. Interestingly the effective diffusion
coefficient becomes negative for some values of the signal phase. However this
does not imply that the periodic driving can be used to concentrate an
ensemble of these systems as the period averaged effective diffusion coefficient $\bar
\kappa^{(2)}$ is always positive.

\section{Conclusion}
We have considered the drift and diffusion behavior of periodic
renewal processes and presented a general theory to express the mean frequency
and effective diffusion constant and also higher order cumulant growth
coefficients in terms of
the underlying time dependent waiting time distribution. The results
of this approach where analytically shown to coincide with known
results in the undriven case. 
For the periodically driven situation we confirmed the results of the
theory by numerical investigations.
We also showed agreement with a different approach for a more
restricted class of renewal processes presented in \cite{prager}.
The mean frequency and effective diffusion coefficient may be used to
quantify stochastic synchronization. Taking into account the large amount of systems whose dynamics can be modeled as renewal
processes, we anticipate a widespread applicability of our results, ranging from
synchronization in neurons to the investigation of transport
properties in molecular motors.

\appendix
\section{Expansion of the probability density governed by a
  Kramers-Moyal equation} \label{herleitung}
Our aim is to express the phase distribution
$\PP(\contvar-\Delta\contvar,t-\tau)$ in terms of $\PP(\contvar,t)$
and its derivatives with respect to $\contvar$, $\partial^n/\partial
\contvar^n \PP(\contvar,t)$.  To this end we start by expanding
\mbox{$\PP(\contvar-\Delta\contvar,t-\tau))$} in a Taylor series around
$\contvar$ and $t$,
\begin{eqnarray*}
\PP(\contvar-\Delta\contvar,t-\tau)=\sum_{n=0}^\infty\sum_{m=0}^\infty
    \frac{(-\Delta\contvar)^n(-\tau)^m}{n!m!}\frac{\partial^{n+m}}{\partial
        \contvar^n\partial t^m}\PP(\contvar,t)
\end{eqnarray*}
To process the time derivatives we use the Kramers-Moyal equation  (\ref{generalfpe})
taking care of the explicit time
dependence of $\kappa^{(i)}(t)$  which leads to
\begin{eqnarray*}
\PP(\contvar-\Delta\contvar,t-\tau)
&=& \PP(\contvar,t)-
\Big[\Delta \contvar+
\sum_{m=1}^\infty
\frac{(-\tau)^m}{m!}
\frac{\partial^{m-1}\kappa^{(1)}(t)}{\partial t^{m-1}}
\Big]\pde{\contvar}\PP(\contvar,t)\\
&&\hspace{-3.5cm}+
\Big[\frac{\Delta \contvar^2}{2}+
\Delta \contvar\sum_{m=1}^\infty
\frac{(-\tau)^m}{m!}\frac{\partial^{m-1}\kappa^{(1)}(t)}{\partial
  t^{m-1}}
+\frac{1}{2}\sum_{m=1}^\infty
\frac{(-\tau)^m}{m!}\frac{\partial^{m-1}\kappa^{(2)}(t)}{\partial t^{m-1}}
\\
&&\hspace{-3.5cm}+\sum_{m=2}^\infty\frac{(-\tau)^m}{m!} \sum_{l=1}^{m-1}
\binom{m-1}{l}
\frac{\partial^{m-1-l}\kappa^{(1)}(t)}{\partial t^{m-1-l}}
\frac{\partial^{l-1}\kappa^{(1)}(t)}{\partial t^{l-1}}
\Big]\pdz{\contvar}\PP(\contvar,t)
+O(3).
\end{eqnarray*}
where $O(3)$ denotes third or higher  derivatives of
$\PP(\contvar,t)$ with respect to $\contvar$. 
The sums containing the coefficients  $\kappa^{(n)}(t)$ in a linear way 
be further evaluated, leading to
\begin{eqnarray*}
\sum_{m=1}^\infty \frac{(-\tau)^m}{m!}\frac{\partial^{m-1}\kappa^{(n)}(t)}{\partial t^{m-1}}
&=&-\sum_{m=0}^\infty
 \frac{1}{m!}\frac{\partial^{m}\kappa^{(n)}(t)}{\partial t^{m}}
 \int_0^\tau d\tau'(-\tau')^{m}\\
&=&-\int_0^\tau d\tau'\kappa^{(n)}(t-\tau')
\end{eqnarray*}
The last term can be simplified to give
\begin{eqnarray*}
\lefteqn{  \sum_{m=2}^\infty\frac{(-\tau)^m}{m!} \sum_{l=1}^{m-1}
\binom{m-1}{l}
\frac{\partial^{m-1-l}\kappa^{(1)}(t)}{\partial t^{m-1-l}}
\frac{\partial^{l-1}\kappa^{(1)}(t)}{\partial t^{l-1}}}\\
&=&\sum_{l=0}^\infty \sum_{m=0}^{\infty} \frac{(-\tau)^{m+l+2}}{(m+l+2)!} 
\binom{m+l+1}{l+1}
\frac{\partial^{m}\kappa^{(1)}(t)}{\partial t^{m}}
\frac{\partial^{l}\kappa^{(1)}(t)}{\partial t^{l}}\\
&=&\int_0^\tau d\tau'
\sum_{m=0}^{\infty}
\frac{(-\tau')^m}{m!}\frac{\partial^{m}\kappa^{(1)}(t)}{\partial t^{m}}
\int_0^{\tau'} d\tau''\sum_{l=0}^\infty
\frac{(-\tau'')^{l}}{l!}\frac{\partial^{l}\kappa^{(1)}(t)}{\partial
  t^{l}}\\ 
&=&\int_0^\tau d\tau' \kappa^{(1)}(t-\tau')
\int_0^{\tau'} d\tau''\kappa^{(1)}(t-\tau'')
\end{eqnarray*}
Thus we eventually arrive at
\begin{eqnarray}\label{fpeexpansion}
\PP(\contvar-\Delta\contvar,t-\tau)
&=&\PP(\contvar,t)+c_t^{(1)}(t-\tau,\Delta\contvar)\pde{\contvar}\PP(\contvar,t)
\nonumber\\&&+c_t^{(2)}(t-\tau,\Delta\contvar)\pdz{\contvar}\PP(\contvar,t)
+O(3)
\end{eqnarray}
where 
\begin{eqnarray}\label{appc1}
c_t^{(1)}(t-\tau,\Delta\contvar)&=&\int_{0}^\tau d\tau' \kappa^{(1)}(t-\tau')-\Delta \contvar
\end{eqnarray}
and
\begin{eqnarray}\label{appc2}
c_t^{(2)}(t-\tau,\Delta\contvar)&=&\frac{\Delta \contvar^2}{2}
-\Delta \contvar \int_{0}^\tau d\tau' \kappa^{(1)}(t-\tau')\\&&\hspace{-2.5cm}
-\frac{1}{2}\int_{0}^\tau d\tau' \kappa^{(2)}(t-\tau')+
\int_{0}^\tau d\tau' \kappa^{(1)}(t-\tau')
\int_0^{\tau'} d\tau''\kappa^{(1)}(t-\tau'')\nonumber
\end{eqnarray}

\section{Relation between moments and cumulants of the number of
  events of a renewal process and its characteristic
functional} \label{relmomcum}
Consider the  moments $M^{(n)}$ defined  by (c.f. eq (\ref{Moments}))
\begin{eqnarray}\label{defmoments}
  M^{(n)}=\pd{n}{z} \exp\big[\sum_{s=1}^\infty G_s (e^z-1)^s\big]\Big|_{z=0}.
\end{eqnarray}
Generally the relation between moments and cumulants is given by
\begin{eqnarray}\label{momcum}
  \sum_{k=0}^\infty \frac{z^k}{k!}M^{(k)}=\exp\big[\sum_{k=1}^\infty \frac{z^k}{k!}K^{(k)}\big].
\end{eqnarray}
From the relation
\begin{eqnarray*}
  M^{(n)}=\pd{n}{z}\sum_{k=0}^\infty \frac{z^k}{k!}M^{(k)}\Big|_{z=0}
\end{eqnarray*}
and eq. (\ref{defmoments}) we deduce (taking into account the
analyticity at $z=0$ of the functions involved) 
\begin{eqnarray*}
  \sum_{k=0}^\infty
  \frac{z^k}{k!}M^{(k)}=\exp\big[\sum_{s=1}^\infty
  G_s (e^z-1)^s\big] 
\end{eqnarray*}
and thus according to eq. (\ref{momcum})
\begin{eqnarray*}
  \sum_{k=1}^\infty \frac{z^k}{k!}K^{(k)}=\sum_{s=1}^\infty
  G_s (e^z-1)^s
\end{eqnarray*}
With
\begin{eqnarray*}
  K^{(n)}=\pd{n}{z}\sum_{k=1}^\infty \frac{z^k}{k!}K^{(k)}\Big|_{z=0}
\end{eqnarray*}
the final result
\begin{eqnarray*}
  K^{(n)}=\pd{n}{z} \sum_{s=1}^\infty G_s (e^z-1)^s\Big|_{z=0}
\end{eqnarray*}
follows.

\section{Relation between the Kramers-Moyal coefficient and the growth
of the cumulants} \label{growthofcummulants}
Consider the stochastic process $\contvar(t)$ governed by
\begin{eqnarray}\label{generalfpeapp}
  \pde{t} \PP(\contvar,t)=\sum_{n=1}^\infty \frac{(-1)^n}{n!} \kappa^{(n)}(t) \pd{n}{\contvar}  \PP(\contvar,t)
\end{eqnarray}
We are interested in the grows of the cumulants $K^{(n)}(t)$ of $\contvar(t)$. The moments 
\begin{eqnarray*}
M^{(n)}(t)=\mean{\contvar^n(t)}=\int_{-\infty}^\infty d\contvar \contvar^n \PP(\contvar,t)
\end{eqnarray*}
obey
\begin{eqnarray*}
  \tde{t}M^{(n)}(t)&=&
\int_{-\infty}^\infty d\contvar \contvar^n \pde{t} \PP(\contvar,t)\\
&=&\sum_{j=1}^{\infty}\frac{(-1)^j}{j!} \kappa^{(j)}(t) \int_{-\infty}^\infty d\contvar \contvar^n
\pd{j}{\contvar}  \PP(\contvar,t)
\end{eqnarray*}
Assuming further that $\PP(\contvar,t)$ decreases sufficiently fast for $\contvar\to \pm
\infty$ such that
\begin{eqnarray*}
  \lim_{\contvar \to \pm \infty }\contvar^n \PP(\contvar,t)=0
\end{eqnarray*}
the above expression can be evaluated using integration by parts to
give 
\begin{eqnarray}\label{momdyn}
  \tde{t}M^{(n)}(t)&=&\sum_{j=1}^{\infty} \frac{\kappa^{(j)}(t)}{j!}
  \int_{-\infty}^\infty d\contvar [\pd{j}{\contvar} \contvar^n]
  \PP(\contvar,t)\nonumber\\
&=&\sum_{j=1}^{n} \frac{\kappa^{(j)}(t)}{j!}
  \int_{-\infty}^\infty d\contvar \frac{n!}{(n-j)!}\contvar^{n-j}
  \PP(\contvar,t)\nonumber\\
&=&\sum_{j=1}^{n} \kappa^{(j)}(t) \binom{n}{j} M^{(n-j)}(t)
\end{eqnarray}
Now the moments and cumulants are related by
\begin{eqnarray}\label{momcum3}
  \sum_{k=0}^\infty \frac{z^k}{k!}M^{(k)}(t)=\exp\big[\sum_{k=1}^\infty \frac{z^k}{k!}K^{(k)}(t)\big].
\end{eqnarray}
and thus by differentiating this equation with respect to $t$
\begin{eqnarray*}
  \sum_{k=0}^\infty \frac{z^k}{k!}\tde{t}M^{(k)}(t)=
\sum_{k=1}^\infty \frac{z^k}{k!}\tde{t}K^{(k)}(t)\sum_{k=0}^\infty \frac{z^k}{k!}M^{(k)}(t).
\end{eqnarray*}
Inserting the moments dynamic eq. (\ref{momdyn}) into the left hand
side of this equation, it can be easily checked that
\begin{eqnarray*}
  \tde{t}K^{(n)}(t)=\kappa^{(n)}(t).
\end{eqnarray*}

\section{The third cumulant growth coefficient $\kappa^{(3)}(t)$}\label{K3}
The third coefficient $\kappa^{(3)}(t)$ can be evaluated by equating the
coefficients corresponding to the order
$\pd{2}{\contvar}\PP(\contvar,t)$ in eq.
(\ref{interstep}). Inserting the expansion
(\ref{fpeexpansion}) into this equation  leads to
\begin{eqnarray}
  \frac{1}{6}\int_0^\infty d\tau
  \kappa^{(3)}(t-\tau)z(\tau,t-\tau)&=&\\
  &&\hspace{-5cm}\frac{1}{2}\int_0^\infty d\tau
  \kappa^{(2)}(t-\tau)\big[\int_{0}^\tau d\tau'\kappa^{(1)}(t-\tau')
-\frac{1}{2}\big]z(\tau,t-\tau)\nonumber\\&&\hspace{-5cm}
-\int_0^\infty d\tau
  \kappa^{(1)}(t-\tau)\big[
\frac{1}{6}
-\frac{1}{2}\int_{0}^\tau d\tau' \kappa^{(1)}(t-\tau')
-\frac{1}{2}\int_{0}^\tau d\tau' \kappa^{(2)}(t-\tau')\nonumber\\&&\hspace{-5cm}+
\int_{0}^\tau d\tau' \kappa^{(1)}(t-\tau')
\int_0^{\tau'} d\tau''\kappa^{(1)}(t-\tau'')
\big]z(\tau,t-\tau)
\label{k3}
\end{eqnarray}
or using eqs. (\ref{omega}) and (\ref{D2})
\begin{eqnarray} \label{k32}
\int_0^\infty d\tau
  \kappa^{(3)}(t-\tau)z(\tau,t-\tau)&=&
\int_0^\infty d\tau z(\tau,t-\tau)\\&&\hspace{-4cm}\nonumber
  \Big[3 \kappa^{(2)}(t-\tau)\int_{0}^\tau d\tau'\kappa^{(1)}(t-\tau')+3\kappa^{(1)}(t-\tau)\int_{0}^\tau
  d\tau'\kappa^{(2)}(t-\tau')\\&&\hspace{-4cm}-6
\kappa^{(1)}(t-\tau)\int_{0}^\tau d\tau'\kappa^{(1)}(t-\tau')\int_{0}^{\tau'}
  d\tau''\kappa^{(1)}(t-\tau'')\Big]+1\nonumber
\end{eqnarray}
For an undriven renewal process eq. (\ref{k32}) can be directly
solved, leading to the time independent coefficient
\begin{eqnarray*}
  \kappa^{(3)}(t)&=&\intrinsicl^3\frac{\mean{T}^4-3\mean{T^2}\mean{T}^2+3\mean{T^2}^2-\mean{T}\mean{T^3}}{\mean{T}^5}
\end{eqnarray*}
For a rate process with periodically modulated rate $\gamma(t)$  (\ref{k32}) eventually leads to
\begin{eqnarray*}
  \kappa^{(3)}(t)=\intrinsicl^3 \gamma(t).
\end{eqnarray*}

\begin{figure}[htbp]
  \centering
 \includegraphics[width=10cm]{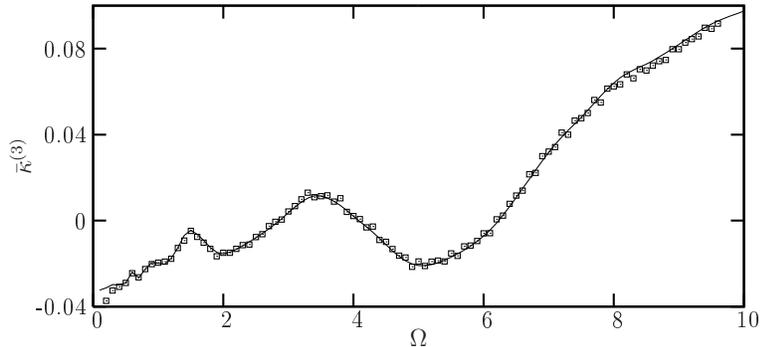} 
\caption{  
\label{comptheosimk3}
Comparison of  $\bar \kappa^{(3)}=\frac{1}{\TT}\int_0^{\TT} dt
\kappa^{(3)}(t)$, where $\TT=2\pi/\Omega$ is the period of the signal,
 for the toy model eq. (\ref{toymodel}) with $T_0=0.5$, $T_1=1.5$ and
 $\gamma=1.7$.
The solid lines are results of the theory eq. (\ref{k32})  numerically evaluated according to
eq. (\ref{k3fourier})  truncated to 20 coefficients,
while the symbols are obtained from direct simulations of
the driven renewal process. Both results agree very well.
}
\end{figure}

Generally, the periodic solution
\begin{eqnarray*}
   \kappa^{(3)}(t)=\sum_{k=-\infty}^\infty \kappa^{(3)}_k \exp(i k \Omega t),\qquad
   \kappa^{(3)}_k=\frac{1}{\TT} \int_0^{\TT} dt \kappa^{(3)}(t) \exp(-i k \Omega t),
\end{eqnarray*}
of eq. (\ref{k32}) can be numerically obtained in Fourier space as a
solution of the (infinite) set of linear equations
\begin{eqnarray}\label{k3fourier}
\lefteqn{  \sum_{k=-\infty}^\infty \kappa^{(3)}_{k} z_{m-k,m}=}\\&&\nonumber
3 \sum_{k=-\infty}^\infty \Big[\big[\sum_{l=-\infty,l\neq 0}^\infty
\frac{\kappa^{(2)}_l \kappa^{(1)}_k+\kappa^{(1)}_l
  \kappa^{(2)}_k}{il\Omega}(z_{m-k-l,m-l}-z_{m-k-l,m})\big]\\&&\hspace{2cm}+\big[\kappa^{(2)}_0 
\kappa^{(1)}_k+\kappa^{(1)}_0
\kappa^{(2)}_k\big]  h_{m-k,m}\Big]\nonumber
\\&&\hspace{-0.3cm}
-6\sum_{k=-\infty}^\infty \Big[\big[\sum_{l=-\infty,l\neq 0}^\infty
\sum_{j=-\infty,j\neq 0,-l}^\infty\nonumber\\&&\hspace{1cm}
\frac{\kappa^{(1)}_k \kappa^{(1)}_l
  \kappa^{(1)}_j}{j\Omega^2}\big(\frac{1}{l}(z_{m-k-l-j,m-j}-z_{m-k-l-j,m-j-l})\nonumber\\&&\hspace{2.5cm}+
\frac{1}{j+l}(z_{m-k-l-j,m-j-l}-z_{m-k-l-j,m})\big)\big]\nonumber\\
&&\hspace{-0.5cm} +\sum_{l=-\infty,l\neq 0}^\infty
\frac{\kappa^{(1)}_k \kappa^{(1)}_l
  \kappa^{(1)}_{-l}}{l^2 \Omega^2}\big(z_{m-k,m}-z_{m-k,m-l}+i l \Omega h_{m-k,m}\big)\nonumber\\
&&
\hspace{-0.5cm}+\sum_{l=-\infty,l\neq 0}^\infty
\frac{\kappa^{(1)}_k \kappa^{(1)}_l
  \kappa^{(1)}_0}{i l\Omega}(h_{m-k-l,m-l}-h_{m-k-l,m})+\kappa^{(1)}_k \kappa^{(1)}_0
  \kappa^{(1)}_0j_{m-k,m}\Big]\nonumber\\&&+\intrinsicl^3 \delta_{m,0}.\nonumber
\end{eqnarray}
where we used the Fourier decompositions of $\kappa^{(1)}$ and
$\kappa^{(2)}$ according to eqs. (\ref{fourierdecompkappa1}) while
$z_{k,l}$ and $h_{k,l}$ are defined in eq. (\ref{fouriercoeffs}) and
$j_{k,l}$ is defined as
\begin{eqnarray*}\label{jj}
  j_{k,l}=\frac{1}{2} \int_0^\infty d\tau \tau^2 z_k(\tau) \exp(-il \Omega \tau).
\end{eqnarray*}
For the toy model introduced in section \ref{toymod}  we have evaluated the period
average of $\kappa^{(3)}(t)$ both according to the theory,
eqs. (\ref{k32}) and (\ref{k3fourier}) and from direct simulations of
the driven renewal process. Both results agree very well
(Fig. (\ref{comptheosimk3})), thus  confirming our theory.

\section{ Equivalence with the two state model for excitable systems}\label{equivts}
In a former work \cite{prager} we investigated stochastic
synchronization in a non Markovian two state model for excitable systems.
Such an excitable dynamics can be approximated by two steps, namely an
excitation from the stable fixed point (rest state), which can be well modeled by a rate process, 
followed by a long excursion in phase space, which constitutes the
excited and refractory state. This excursion takes some time which is
distributed according to a waiting time distribution. As the
excitation step is much more sensible to the periodic driving than the
time spent on the excitation loop, we may assume this waiting time
distribution $\tilde w(\tau)$ to be independent on the running time.
The rate $\gamma(t)$ of the excitation step however, depends on the
periodic signal and thus periodically on the time $t$.

In this model we have  defined  a phase in order to investigate
synchronization between the signal and the system's dynamics. This
phase was chosen to increase by $2\pi$ each time the system has been
excited and returned back to the rest state. We have derived
expressions for the mean phase velocity $\omega(t)$ and effective phase
diffusion constant $D(t)$. In the following we show that these
expressions agree with the corresponding expressions of our
general theory presented here.
Namely by choosing $\intrinsicl=2\pi$ 
the phase velocity $\omega(t)$ can be identified with $\kappa^{(1)}(t)$
while the effective phase diffusion coefficient $D(t)$ in this prior
work differed from our definition of $\kappa^{(2)}(t)$ by a factor of 2,
i.e. $\kappa^{(2)}(t)\equiv 2 D(t)$.

In this two state model the mean phase velocity was given by
\begin{eqnarray}\label{relomegaq2}
  \omega(t)=2 \pi \gamma(t) q_2^{(0)}(t)
\end{eqnarray}
where the dynamics of $q_2^{(0)}(t)$ are governed by
\begin{eqnarray}\label{tsmodel}
  1-q_2^{(0)}(t)=\int_0^\infty d\tau \gamma(t-\tau) q_2^{(0)}(t-\tau)
  \tilde z(\tau) 
\end{eqnarray}
The effective diffusion coefficient was calculated as 
\begin{eqnarray}\label{ansatzD}
  D(t)=-2 \pi \gamma(t) q_2^{(1)}(t)+\pi \omega(t)
\end{eqnarray}
with
\begin{eqnarray}\label{dynq2}
  -2\pi q_2^{(1)}(t)&=&2\pi \int_0^\infty d\tau \tilde
  z(\tau)\gamma(t-\tau)q_2^{(1)}(t-\tau) 
\\&&+\int_0^\infty d\tau \tilde z(\tau)  \omega(t-\tau)(\int_0^\tau d\tau'
\omega(t-\tau')-2\pi).\nonumber
\end{eqnarray}
The auxiliary variables $q_2^{(0)}$ and $q_2^{(1)}$ can be eliminated
leading to
\begin{eqnarray}\label{tsomega}
  \gamma(t) \int_0^\infty d\tau  \omega(t-\tau) \tilde z(\tau)+\omega(t)=2\pi \gamma(t)
\end{eqnarray}
and
\begin{eqnarray}\label{tsD}
\lefteqn{  \gamma(t)\int_0^\infty d\tau  D(t-\tau) \tilde z(\tau)+D(t)=} \\&&
\hspace{1cm}-2\pi^2
  \gamma(t)+2 \pi \omega(t)
+\gamma(t)\int_0^\infty d\tau \omega(t-\tau)\int_0^\tau
  d\tau' \omega(t-\tau') \tilde z(\tau).\nonumber
\end{eqnarray}
In the following we show that these equations for $\omega(t)$ and
$D(t)$ are equivalent to our general eqs. (\ref{omega}) and (\ref{D2}) which
for $\omega(t)$ and $D(t)$ read
\begin{eqnarray}
 \int_0^\infty d\tau \omega(t-\tau)z(\tau,t-\tau)&=&2\pi\label{omegaex}\\
\end{eqnarray}
and
\begin{eqnarray}
\label{Dex}
\lefteqn{ \int_0^\infty d\tau D(t-\tau)z(\tau,t-\tau)=}\\&&\hspace{1cm}\int_0^\infty d\tau
  \omega(t-\tau)\int_{0}^\tau d\tau'
  \omega(t-\tau+\tau')z(\tau,t-\tau)-2\pi^2\nonumber
\end{eqnarray}
The waiting time between two events is given by the sum of the
excitation time and the time spent on the excitation loop.
The  time dependent waiting time distribution for
one cycle is therefore given by  
\begin{eqnarray}\label{wexcitable}
  w(\tau,t)&=&\int_0^\tau d\tau' \tilde w(\tau') \gamma(t+\tau) \exp(-\int_{t+\tau'}^{t+\tau}
  d\tau'' \gamma(\tau''))\nonumber.
\\&&
\end{eqnarray}
while the corresponding survival probability reads
\begin{eqnarray}\label{zexcitable}
  z(\tau,t)=\int_0^\tau d\tau' \tilde w(\tau') \exp(-\int_{t+\tau'}^{t+\tau}
  d\tau'' \gamma(\tau''))+\tilde z(\tau)
\end{eqnarray}
where $\tilde z(\tau)=1-\int_0^\tau d\tau' \tilde w(\tau)$ is the
probability to spent a time longer than $\tau$ on the excitation loop.
From these equations we obtain the relation 
\begin{eqnarray}\label{relzwexcitable}
  \gamma(t) z(\tau,t-\tau)=w(\tau,t-\tau)+\tilde z(\tau).
\end{eqnarray}

Multiplying eqs. (\ref{omegaex}) and (\ref{Dex}) by $\gamma(t)$ and replacing the time dependent
survival probability $z(\tau,t-\tau)$ according to eq. (\ref{relzwexcitable})  we obtain
\begin{eqnarray}
   \gamma(t)\int_0^\infty d\tau \omega(t-\tau)\tilde z(\tau)+
   \int_0^\infty d\tau \omega(t-\tau)
   w(\tau,t-\tau)&=&2\pi\gamma(t)\label{isomega}
\end{eqnarray}
and
\begin{eqnarray}
   \hspace{-1cm}\gamma(t)\int_0^\infty d\tau 
\big[D(t-\tau)-\omega(t-\tau)\int_0^\tau d\tau' \omega(t-\tau-\tau') \big]
\tilde z(\tau)+\label{isD}\\
   \int_0^\infty d\tau \big[D(t-\tau)-\omega(t-\tau)\int_0^\tau d\tau' \omega(t-\tau-\tau') 
   \big]w(\tau,t-\tau)=-2\pi^2\gamma(t)\nonumber
\end{eqnarray}
The second term on the left hand side in both equations can be further
simplified. To this end we differentiate  eqs. (\ref{omegaex}) and (\ref{Dex})  with respect to $t$ (see appendix
\ref{diffgenmod}), using eq. (\ref{omegaex}) to simplify the time
derivative of eq. (\ref{Dex}) which results in 
\begin{eqnarray}\label{omegaderiv}
\omega(t)=\int_0^\infty d\tau \omega(t-\tau) w(\tau,t-\tau)
\end{eqnarray}
and
\begin{eqnarray}\label{Dderiv}
D(t)=\int_0^\infty \!\!\!d\tau 
\big[D(t-\tau)-\omega(t-\tau)\int_0^\tau \!\!\! d\tau'
\omega(t-\tau-\tau') \big]w(\tau,t-\tau)+2\pi \omega(t).\nonumber \\&&
\end{eqnarray}
Note that the above  equations can be equivalently
derived from  eq. (\ref{dyneqasymp}) by performing the same
procedure we have applied to eq. (\ref{relprobfluxasymp})
in order to obtain eqs. (\ref{omega}) and (\ref{D2}).

Inserting finally
eqs. (\ref{omegaderiv}) and (\ref{Dderiv}) into eqs. (\ref{isomega})
and (\ref{isD}), eventually leads to the two state model eqs. (\ref{tsomega}) and (\ref{tsD}).

Thus we have shown that solutions of $D(t)$ and $\omega(t)$ of the
general theory are also solutions of the two state model for excitable
systems presented in \cite{prager}. As these solutions are in general
unique we have shown the equivalence between the two models. In
\cite{prager} these solutions were compared with simulations of the
renewal process, showing total agreement within simulation precision.

\section{Time derivatives of integrals involving time dependent
  survival probabilities}\label{diffgenmod}
Let $w(\tau,t)$ be a time dependent waiting time distribution, i.e.
\begin{eqnarray*}
  w(\tau,z)\ge 0 \quad \forall \; \tau,t \qquad \text{and}\qquad
  \int_0^\infty d\tau w(\tau,t)=1 \quad \forall \;\tau.
\end{eqnarray*}
Let further 
\begin{eqnarray*}
  z(\tau,t)=1-\int_0^\tau d\tau' w(\tau',t)
\end{eqnarray*}
be the corresponding survival probability.
Then for any sufficiently well behaved function $g(t)$ we have
\begin{eqnarray}\label{relzw}
 \tde{t} \int_0^\infty d\tau g(t-\tau) z(\tau,t-\tau)= g(t)-\int_0^\infty d\tau g(t-\tau) w(\tau,t-\tau).
\end{eqnarray}
This can be seen as follows:
\begin{eqnarray*}
\lefteqn{\tde{t} \int_0^\infty d\tau g(t-\tau) z(\tau,t-\tau)}\nonumber\\
&=& \int_0^\infty d\tau \Big(-\tde{\tau} g(t-\tau)
    z(\tau,t-\tau)+g(t-\tau)\tde{\tau'}
    z(\tau',t-\tau)\big|_{\tau'=\tau}\Big)\nonumber\\
&=&-g(t-\tau)
    z(\tau,t-\tau)\Big|_{\tau=0}^\infty-\int_0^\infty d\tau g(t-\tau)
    w(\tau,t-\tau)\nonumber\\
&=&g(t)-\int_0^\infty d\tau g(t-\tau) w(\tau,t-\tau).
\end{eqnarray*}
where in the last step we assumed that $g(t-\tau)z(\tau,t-\tau)$ converges
to zero as $\tau\to \infty$.

We further have the identity
\begin{eqnarray}\label{relzw2}
\lefteqn{  \tde{t}\int_0^\infty d\tau
  g(t-\tau)\int_{0}^\tau d\tau'
  g(t-\tau+\tau')z(\tau,t-\tau)
}\nonumber\\&=&\int_0^\infty d\tau
  \Big(-\tde{\tau }\Big[g(t-\tau)\int_{0}^\tau d\tau'
  g(t-\tau+\tau')z(\tau,t-\tau)\Big]\nonumber\\&&\quad
+g(t-\tau)g(t)z(\tau,t-\tau)
-  g(t-\tau)\int_{0}^\tau d\tau'
  g(t-\tau+\tau')w(\tau,t-\tau)\Big)
\nonumber\\&=&
g(t)\int_0^\infty d\tau g(t-\tau)z(\tau,t-\tau)
\\&&\hspace{2cm}- \int_0^\infty d\tau g(t-\tau)\int_{0}^\tau d\tau'
  g(t-\tau+\tau')w(\tau,t-\tau)\nonumber.
\end{eqnarray}

\end{document}